\documentclass{moriondNOTITLES}

\usepackage{amsmath,amssymb,mathrsfs}
\usepackage[rightcaption]{sidecap}
\sidecaptionvpos{figure}{c}

\def \msun  {\rm{M}_\odot}

\newcommand{\Photo}{}

\begin{document}
\vspace*{4cm}
\title{Tests of General Relativity with GW230529: \\ a neutron star merging with a lower mass-gap compact object}

\author{Elise M. S\"anger}

\address{Max Planck Institute for Gravitational Physics (Albert Einstein Institute), \\ Am M{\"u}hlenberg 1, 14476 Potsdam, Germany}

\maketitle
\abstracts{
We performed tests of General Relativity on gravitational wave signal GW230529\_181500, which comes from what is most likely a neutrons star merging with a black hole in the lower mass gap. We used two different frameworks to perform parameterized inspiral tests. We find that the signal is consistent with General Relativity for all deviation parameters and we obtain particularly tight constraints on dipole radiation. We discuss some challenges that arise when analyzing this signal, namely biases due to correlations with tidal effects and the degeneracy between the deviation parameter at Newtonian order and the chirp mass. We also performed a theory-specific test for Einstein-scalar-Gauss-Bonnet gravity where we obtain the best constraints on this theory to date.
}

\section{Introduction}

On 29 May 2023, the LIGO Livingston observatory~\cite{LIGOScientific:2014pky,Capote:2024rmo} detected the gravitational-wave (GW) signal GW230529\_181500 (hereafter GW230529 for briefty) from the merger of what is most likely a neutron star with a black hole in the lower mass-gap~\cite{LIGOScientific:2024elc}. This signal had a signal-to-noise ratio of 11.6 and a false-alarm rate of less than 1 in 1000 years. Follow-up analyses estimated the source component masses to be $\ensuremath{3.6_{-1.1}^{+0.8}}~\msun$ and $\ensuremath{1.4_{-0.2}^{+0.6}}~\msun$ (90\% credible interval), which puts the primary squarely in the lower mass gap of $\sim 2 \textup{ -- } 5~\msun$. 

GW230529, with the compact object within the hypothesized mass gap, provides an opportunity to test General Relativity (GR) in a region of parameter space previously unexplored by strong-field tests of GR~\cite{LIGOScientific:2020tif,LIGOScientific:2021sio}. Because of its long inspiral signal, it is especially of interest for tests of GR looking for deviations in the inspiral regime. We therefore performed theory agnostic, parameterized inspiral tests of GR using two different frameworks: the Flexible Theory Independent (FTI) method~\cite{Mehta:2022pcn} and the Test Infrastructure for General Relativity (TIGER)~\cite{Roy:2025gzv}. To demonstrate the importance of this signal for testing GR, we also performed a theory specific test for Einstein-scalar-Gauss-Bonnet, an alternative theory of gravity.

\section{Parameterized inspiral tests of GR}

The post-Newtonian (PN) formalism can be used to approximate the GW signal from the early inspiral of a compact binary~\cite{Blanchet:2013haa}. The PN approximation expands the waveform in powers of the velocity $v$, where $\mathcal{O}(v^{2n})$ relative to leading order is refered to as the $n$PN order. In GR, the frequency domain phase is then given by
\begin{equation}
\Psi_{\ell m}^\text{GR}(f) = 2\pi f t_c - \phi_c - \frac{\pi}{4} + \frac{3}{128\eta v^5}\frac{m}{2}  \sum_{n=0}^7\left( \psi_n^\text{GR} + \psi_{n(l)}^\text{GR} \log v \right) v^n,
\end{equation}
where $f$ is the GW frequency, $t_c, \phi_c$ are the time and phase at coalescence, $v = (2\pi fM/m)^{1/3}$, $M=m_1+m_2$ is the detector frame total mass, and $\eta=m_1 m_2/M^2$ is the symmetric mass ratio. The $(n/2)$PN coefficients in GR $\psi_n^\text{GR}$ and $\psi_{n(l)}^\text{GR}$ depend only on the intrinsic parameters of the binary. The GW signal can be descomposed into spherical harmonics and the subscript $\ell m$ denotes the $(\ell,m)$-mode in the mode decomposition.

In beyond-GR theories, these PN coefficients can be different from the ones in GR. We can test for deviations from GR during the inspiral by introducing deviation parameters $\delta\hat{\varphi}_n$ and $\delta\hat{\varphi}_{n(l)}$ that are fractional deviations of the corresponding PN coefficients in GR. We introduce absolute deviations instead when the $(n/2)$PN coefficient vanishes in GR (i.e. for $n=-2,1$). This gives a correction to the frequency domain phase of the form
\begin{equation}
\delta \Psi_{\ell m}(f) = \frac{3}{128\eta v^5}\frac{m}{2} \left( \sum_{n=-2}^7 \delta\hat{\varphi}_n \psi_n^\mathrm{GR} v^n + \sum_{n=5}^6 \delta\hat{\varphi}_{n(l)} \psi_{n(l)}^\mathrm{GR} v^n \log v \right).
\end{equation}
GR is recovered in the limit $\delta\hat{\varphi}_{n}, \delta\hat{\varphi}_{n(l)} \rightarrow 0$. The corrections are only applied to the inspiral portion of the signal and the merger-ringdown is left the same as in GR.

We use two different frameworks that can perform this type of test
of GR: FTI~\cite{Mehta:2022pcn} and TIGER~\cite{Roy:2025gzv}. These two differ in their exact implementation of the inspiral test and in the GR waveform models used as a baseline. FTI is only available for aligned spin waveforms so we use the \textsc{SEOBNRv4} waveform family, while TIGER is specific to the \textsc{IMRPhenomX} waveform family and can use precessing spin waveforms. The main difference in the implementation occurs at the transition from a non-GR inspiral to the GR merger ringdown.

These parameterized inspiral tests are typically performed allowing only one deviation parameter to vary at a time. This means the analysis is repeated for each deviation parameter.

\section{Results}

\begin{figure}
\centering
\includegraphics[width=\textwidth]{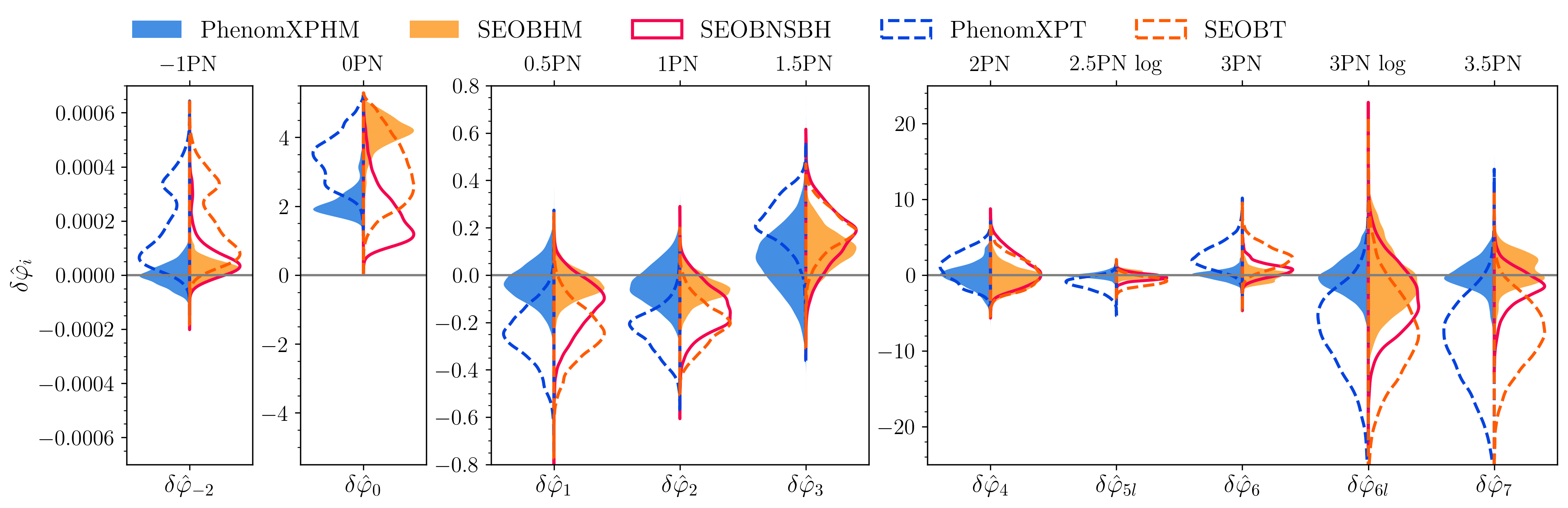}
\caption{The posterior distributions for the different deviation parameters $\delta\hat{\varphi}_n, \delta\hat{\varphi}_{n(l)}$ for GW230529. The blue histograms are obtained with TIGER using the \textsc{IMRPhenomX} waveform family. The orange posteriors are results from FTI using the \textsc{SEOBNRv4} waveform family. The filled violins are for binary black hole waveform models, while the dashed lines are binary neutron star models which include tidal effects for both compact objects, and the red solid line is a neutron star--black hole model with tides only on the secondary. GR is recovered at $\delta\hat{\varphi}_i=0$, which is indicated by the horizontal gray line.}
\label{fig:violin-plot}
\end{figure}

The posteriors for the different deviation parameters are shown in Fig.~\ref{fig:violin-plot}. Comparing the results of both tests, we see that they are consistent with each other and that all results are consistent with GR except for the 0PN results. We will discuss the 0PN results in more detail below.

The bounds obtained for the -1PN, 0.5PN, and 1PN deviation parameters are some of the best to date. The bound obtained for GW230529 on dipole radiation (-1PN) of $|\delta\hat{\varphi}_{-2}| \lesssim 8 \times 10^{-5}$ is an order of magnitude tighter than the combined bounds from GWTC-3~\cite{LIGOScientific:2021sio}. The only tighter bound obtained with GWs is for the binary neutron star GW170817.

\begin{SCfigure}
\centering
\includegraphics[width=0.5\textwidth]{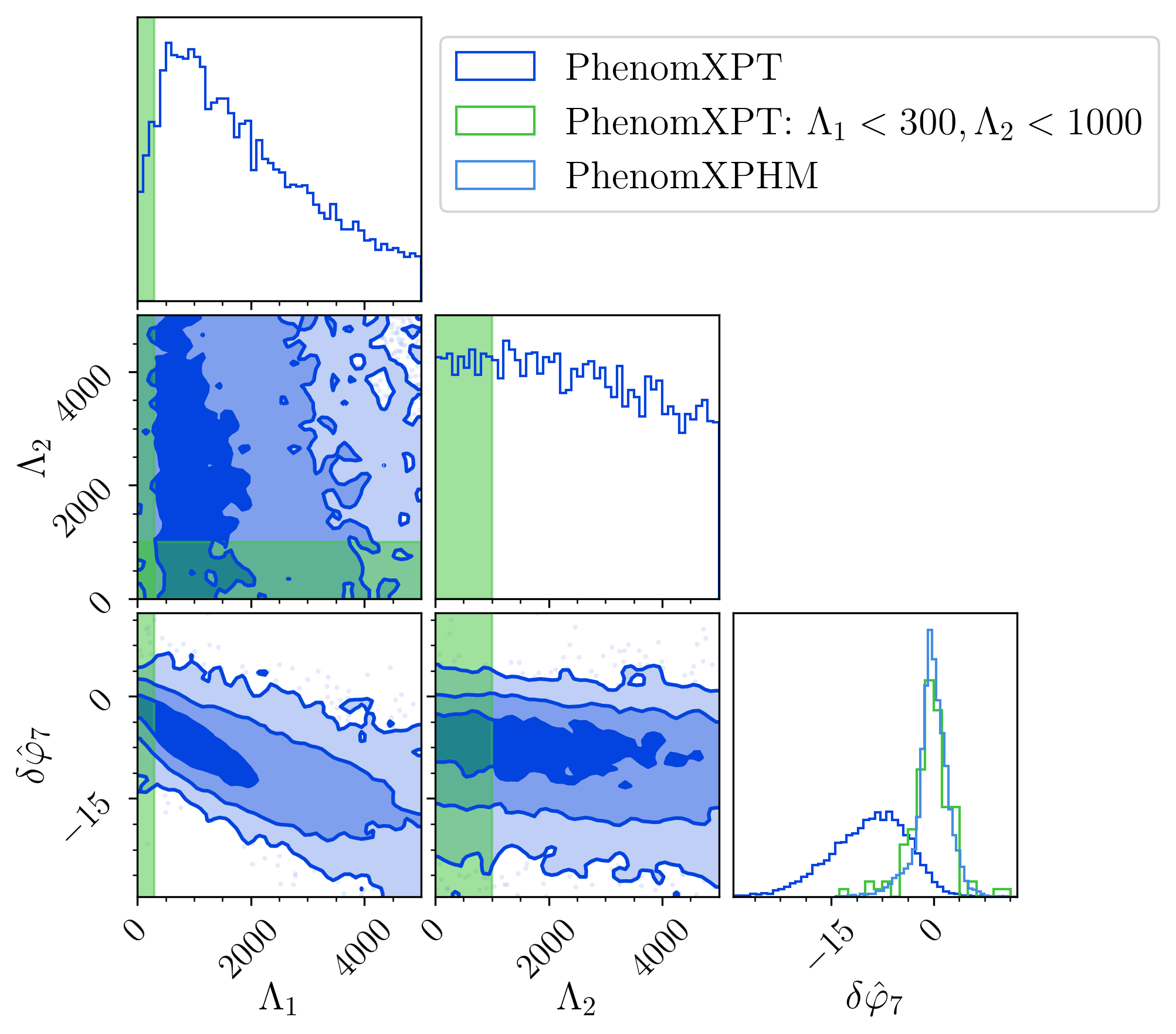}
\caption{The 2D posteriors between the tidal deformabilities $\Lambda_i$ and deviation parameters $\delta\hat{\varphi}_7$ (dark blue) obtained using a binary neutron star model. The green shaded regions are more realistic ranges for the tides, which, when restricting the posterior to those regions, give the green histogram for the deviation parameter. The light blue histogram is the posterior obtained with a binary black hole model. We see that there is a correlation between the tides and the deviation parameter, which leads to a shift away from GR. When restricting the tides to more reasonable values, this shift disappears.}
\label{fig:tides}
\end{SCfigure}

\textbf{\textit{Tidal effects~}} The source of GW230529 is most likely contains a neutron star, so we also performed analyses with waveforms that include tidal effects. The deviation-parameter posteriors obtained with these waveforms are broader and shifted away from GR compared to the results obtained with binary black hole models. In Fig.~\ref{fig:tides}, we see that this is due to correlations between the deviation parameters and tidal deformabilities $\Lambda_i$. For GW230529, the tidal effects are not well constrained and even allow for unrealistically high values~\cite{LIGOScientific:2024elc}, leading to the broad and shifted posteriors for $\delta\hat{\varphi}_i$. By constraining the tides to the more realistic values of $\Lambda_1 < 300$ and $\Lambda_2 < 1000$, the shift and broadening (mostly) disappear (see green histogram in Fig.~\ref{fig:tides}). We therefore use the results from the BBH waveform models as representative.

\begin{SCfigure}[][h]
\centering
\includegraphics[width=0.5\textwidth]{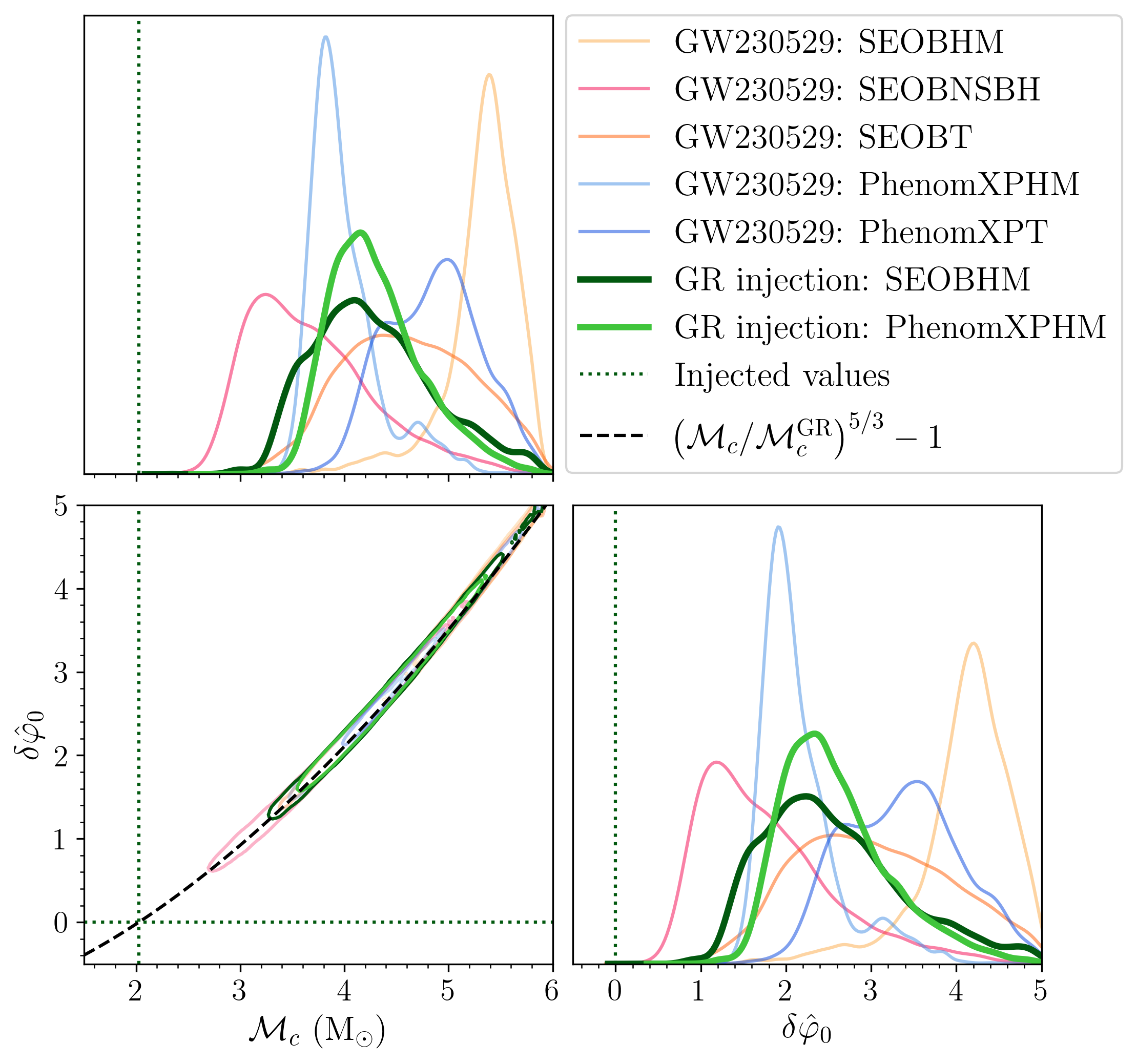} 
\caption{The posteriors for the chirp mass $\mathcal{M}_c$ and 0PN deviation parameter $\delta\hat{\varphi}_0$ for zero-noise injections of the maximum likelihood waveform from the corresponding GR run (green), compared to the posteriors obtained for GW230529 with the different waveform models (other colors). We see that the results from the GR injections are biased towards higher $\mathcal{M}_c$ and positive $\delta\hat{\varphi}_0$ and are similar to the GW230529 results. The black dashed line is the expected degeneracy between $\mathcal{M}_c$ and $\delta\hat{\varphi}_0$ based on the GR maximum likelihood chirp mass $\mathcal{M}_c^\mathrm{GR}$ (green dotted line). We see that the posteriors follow the expected correlation.}
\label{fig:0pn}
\end{SCfigure}

\textbf{\textit{0PN degeneracy~}} When approximating the GW phase to Newtonian order, there is a degeneracy between the chirp mass and the 0PN deviation parameter. Assuming that the binary is consistent with GR and that the true chirp mass is equal to $\mathcal{M}_c^{\mathrm{GR}}$, this degeneracy is given by~\cite{Payne:2023kwj}
$\delta\hat{\varphi}_0 = \left( \mathcal{M}_c / \mathcal{M}_c^{\mathrm{GR}} \right)^{5/3} -1$. 
This means that the chirp mass and $\delta\hat{\varphi}_0$ can move along this degeneracy without significantly changing the waveform. For higher mass systems, this degeneracy is broken by higher PN contributions in the late inspiral and by the merger-ringdown. For GW230529, however, the merger-ringdown is outside the analyzed frequency band, so there is a strong correlation between $\mathcal{M}_c$ and $\delta\hat{\varphi}_0$. This can be seen in the results shown in Fig.~\ref{fig:0pn}, where we also notice that the posterior is shifted away from GR and towards higher chirp mass. Since this shift is also observed in a zero-noise GR injection (green in Fig.~\ref{fig:0pn}), it is likely a false deviation from GR. The cause of this shift is probably a combination of sampling issues due to a sharply peaked likelihood, the choice of priors, the event being single detector, and possibly noise features.

\section{Constraining Einstein-scalar-Gauss-Bonnet}

To illustrate the importance of GW230529 for testing GR, we put constraints on Einstein-scalar-Gauss-Bonnet (ESGB), which is a modified gravity theory where a scalar field is coupled to the Gauss-Bonnet density. The leading order correction in ESGB appears at $-1$PN due to scalar dipole radiation and is given by~\cite{Lyu:2022gdr}
\begin{equation}
\delta\hat{\varphi}_{-2} = -5\ell_{\mathrm{GB}}^4 \frac{\left(m_1^2 s_2 - m_2^2 s_1\right)^2}{168 m_1^4 m_2^4},
\end{equation}
where $\ell_{\mathrm{GB}}$ is the coupling strength of the theory and $s_i$ is the scalar charge of the compact object. We can use this to map the posteriors obtained for $\delta\hat{\varphi}_{-2}$ to the ESGB coupling $\ell_{\mathrm{GB}}$. This gives us an upper bound on the ESGB coupling of $\ell_{\mathrm{GB}} \lesssim 0.67~\msun$. We have also performed a theory-specific test for ESGB gravity by implementing all corrections up to 1.5PN in the FTI framework~\cite{Sanger:2024axs}. This allows us to sample directly over the ESGB coupling $\ell_{\mathrm{GB}}$ and we obtain a slightly better bound of $\ell_{\mathrm{GB}} \lesssim 0.51~\msun$, which is the best bound on the ESGB coupling to date.

\section{Conclusion}

We found no evidence for deviations from GR for GW230529. We obtained particularly tight constraints on deviations from GR at low-PN orders and on ESGB gravity. This demonstrates the importance of signals similar to GW230529 for testing GR. We also discussed some challenges that arise when analyzing signals from low-mass systems such as GW230529, namely tidal effects and the degeneracy between the 0PN deviation parameter and the chirp mass. Hopefully, future, more detailed studies of these effects will lead to ways to better account for, and perhaps even mitigate, the systematic biases they introduce.

\section*{Acknowledgments}

I would like to thank all authors of the original paper~\cite{Sanger:2024axs} for their contributions. I would especially like to thank Soumen Roy for performing all the TIGER analyses and for many fruitful discussions. I would also like to thank the organizers of the \textit{59th Recontres de Moriond} for giving me the opportunity to present this work and for organizing an amazing conference.
This material is based upon work supported by NSF's LIGO Laboratory which is a major facility fully funded by the National Science Foundation. 

\small 

\section*{References}

\bibliography{elise_sanger.bib}

\end{document}